\documentclass[journal]{IEEEtran}
\usepackage{cite}
\usepackage{graphicx}
\usepackage{amsmath}
\usepackage{amssymb}
\usepackage{fancyhdr}
\newlength{\zeroheight}
\usepackage{float}
\usepackage{epstopdf}

\begin{document}

\title{\LARGE \bf  Monitoring voltage collapse margin with synchrophasors across transmission corridors with multiple lines and multiple contingencies}
\author{
\IEEEauthorblockN{ Lina Ramirez\hspace{2cm}Ian Dobson}\\
\IEEEauthorblockA{Electrical and Computer Engineering Dept.\\
Iowa State University\\
Ames IA USA \\
linar@iastate.edu, dobson@iastate.edu}}
%\fancyhead[c]{\textnormal{\small submitted to IEEE Power and Energy Society General Meeting, July 2015,  Denver CO USA}}
\renewcommand{\headrulewidth}{ 0.0pt}
\fancyfoot[C]{\fontfamily{ptm}\selectfont\fontsize{10}{10}\null}

\maketitle
\thispagestyle{fancy}
\begin{abstract}  
We use synchrophasor measurements of the complex voltage and current at both ends of multiple transmission lines that connect areas of a power system to monitor the online voltage collapse margin. A new reduction is used to reduce the multiple transmission lines to a single line equivalent and determine how to combine the synchrophasor measurements.
Generator reactive power limits can be accommodated.
The results show that this methodology can capture the effect of multiple contingencies inside the transmission corridors, giving awareness to the operators about the severity of contingencies with respect to voltage stability.
\\
\end{abstract}

\begin{IEEEkeywords} Area angle, area voltage, contingency analysis, maximum loadability, phasor measurement units, power system security, smart grid, Th\'evenin equivalent, voltage stability.
\end{IEEEkeywords}

\section*{Notation}
\vspace{2pt}
 \begin{tabular}{ @{}ll @{}}
  $Y_{ij}$&Admittance of line between buses $i$ and $j$\\
 $V_i$&Complex voltage at bus  $i$\\
 $I_{i}$&Complex current injected at bus  $i$\\
  $S_{i}$&Complex power injected at bus  $i$\\
 $V_{ij}$&Complex voltage in line  between buses $i$ and $j$\\
 $I_{ij}$&Complex current in line between buses $i$ and $j$\\
  $g$&Generation bus\\
 $\ell$&Load bus\\
  PMU& Phasor measurement unit\\
 Corridor& Transmission lines that connect areas\\

\end{tabular}

\nocite{*}

\nocite{515191, 387897, 486117, 826465, 1033089, 1397706, 1372806,1425595, 4410569, 4809089, 5442743, 5376473}

\section{Introduction}
\label{intro}

As the load  is concentrated in some areas, and abundant generation is generally distant from the load, some areas export or import bulk  power through transmission corridors.  Contingencies and large transfers of power through the corridors  increase the risk of voltage collapse and blackout. For these reasons, it is useful to monitor the margin to voltage collapse from measurements, so that the operator can take prompt action to restore the margin if it becomes too small.

Of course, voltage collapse under n-1 contingency can be assessed  based on power flow analysis and the state estimator  \cite{5528820}-\cite{6156658}, but there is scope for quickly monitoring multiple outages  based on synchrophasor measurements. Multiple outages are prone to occur during bad weather or cyber-physical attacks.  Nowadays, for reasons of cost  and computational time, multiple contingencies cannot be feasibly evaluated pre-contingency in a systematic way using power flow or the state estimator \cite{Varshney13}. 

In this paper, we address online operational advice about the voltage margin for multiple contingencies  in transmission corridors that are connecting areas. Namely, we propose a complementary methodology for evaluating online the contingencies that are not covered under  pre-contingency n-1  analysis, giving awareness and recommending control center actions to  remedy the problems of voltage stability. 

Since the late nineties, researchers have made vigorous efforts to develop methods based on  synchrophasor measurements to detect voltage stability problems in real time \cite{Glavic11}. However, these approaches are based on a corridor with a single line, and there are difficulties in applying the methods to corridors with multiple transmission lines. Previously in   \cite{RamirezGM14} we addressed this difficulty by proposing a way to combine synchrophasor measurements using the area angle, that approximately reduces a transmission corridor with multiple lines to an equivalent single line, so that the known methods for measuring voltage stability margin for a single line could be applied. This reduction using the  area angle does not yet accommodate  generator reactive power limits, and our initial experience is that the different reduction used in  this paper can give a more accurate  estimate of the margin.
We continue to explore ways to handle the problem of multiple corridors and lines in this paper because it is a key barrier to applying synchrophasor measurements to avoid voltage collapse in practical power transmission systems that often have multiple corridors joining generation and load areas.

Our new methodology reduces multiple lines of transmission corridors to a single line using synchrophasor measurements of  complex power and current at each end of each line in the transmission corridor.    The multiple transmission corridors are reduced to a single line 
while preserving the complex power and currents entering and leaving the corridors, and this reduction shows how the 
synchrophasor measurements should be combined. Then the  known methods for measuring voltage stability margin for a single line can be applied to the combined synchrophasor measurements. This paper presents the complex power reduction and demonstrates  how the combined synchrophasor measurements  can be used to 
measure the voltage collapse margin online under unusual increments of load, line outages, or generation outages. 
We note that a reduction methodology preserving current and complex power was also previously used in the context of 
reduced dynamic models for oscillations in \cite{Chakrabortty11}.

The paper is structured as follows. Section II reviews previous work. Section III shows how to combine the synchrophasor measurements and reduce the system to a single line including reactive limits. Results for the WSCC 9-bus test system are presented in Section IV, and Section V concludes the paper.

\section{Previous work}

\subsection{Contingency analysis}

This subsection reviews some of the  extensive previous work on off-line n-1 contingency screening for voltage collapse, see \cite{5528820}- \cite{6156658}, and references therein. 

Reference \cite{EjebePS88} describes the first  procedure for finding the buses with potential voltage stability problems under contingencies, which is based on power flow and the modifications in reactive power consumption of the remaining system.

An efficient methodology for evaluating the change of the voltage collapse margin under contingencies is indicated in \cite{GreenePS99}. This method is based on the current operating point and the pattern of load increase determined by the load forecast. 
The bifurcation and the  load margin are computed and the load margin sensitivity to line outages is evaluated.  
Reference \cite{FlueckPS02} improves this sensitivity method by using a continuation power flow decreasing the admittance of the outaged line. A new index is proposed that includes the maximum load flow of the remaining lines and  the sensitivities of the load margin under line outages.
The method is further improved in  \cite{FlueckPS04} by using a two parameter continuation power flow, where one parameter is the load increase pattern and the other parameter controls the line outage. In addition, this method includes reactive limits of the generators giving more realistic and accurate results.

The approach that we present in this paper differs from and is complementary to   \cite{5528820}-\cite{6156658} in using post contingency measurements online rather than pre contingency calculations based on the state  estimator.  Our measurement-based approach is less accurate than state estimator methods, but  can work independently of the state estimator,  is fast, and handles multiple contingencies more easily. 

\subsection{Measuring voltage collapse margin  with synchrophasors}

For monitoring voltage stability online, many researchers have made vigorous efforts to apply synchrophasor measurements  \cite{Glavic11}. However, the initial approaches were based on a corridor with a single line, and there are difficulties in directly applying the methods to corridors with multiple transmission lines.  It can be noted that applying the single line methods by increasing one load while the other loads remain constant is an obviously unrealistic system stress. 

Due to these problems, some researchers recognized the importance of developing an online voltage stability tool for a system with multiple transmission lines \cite{WarlandPSC02}, \cite{LarssonBPT03},  \cite{Mingsong08}. However, those approaches require strong assumptions, such as known admittance between the loads or generators, making it impossible to capture changes on the transmission corridor, or assuming known complex voltage in the generator. These assumptions can generate inaccurate results, especially during multiple contingencies. In order to help solve these problems, our previous work   in \cite{RamirezGM14} proposed an initial way to combine synchrophasor measurements based on area angle that approximately reduces a corridor with multiple lines to an equivalent single line, giving a more justifiable and accurate indication of the margin to voltage collapse.

One general problem with the aforementioned methodologies is the reactive power limits of the generators. When a generator bus that is considered as PV changes to PQ, the maximum transfer of power through the transmission corridor is reduced substantially.  Some previous approaches for including reactive limits that are different than our methodology for including reactive limits are in \cite{Glavic11,Milosevic03,DuongEPECS13}.

\section{Reduction of transmission corridor with multiple lines to a single line system using complex powers}

\begin{figure}[]
\centering
\includegraphics [width=0.45\textwidth]{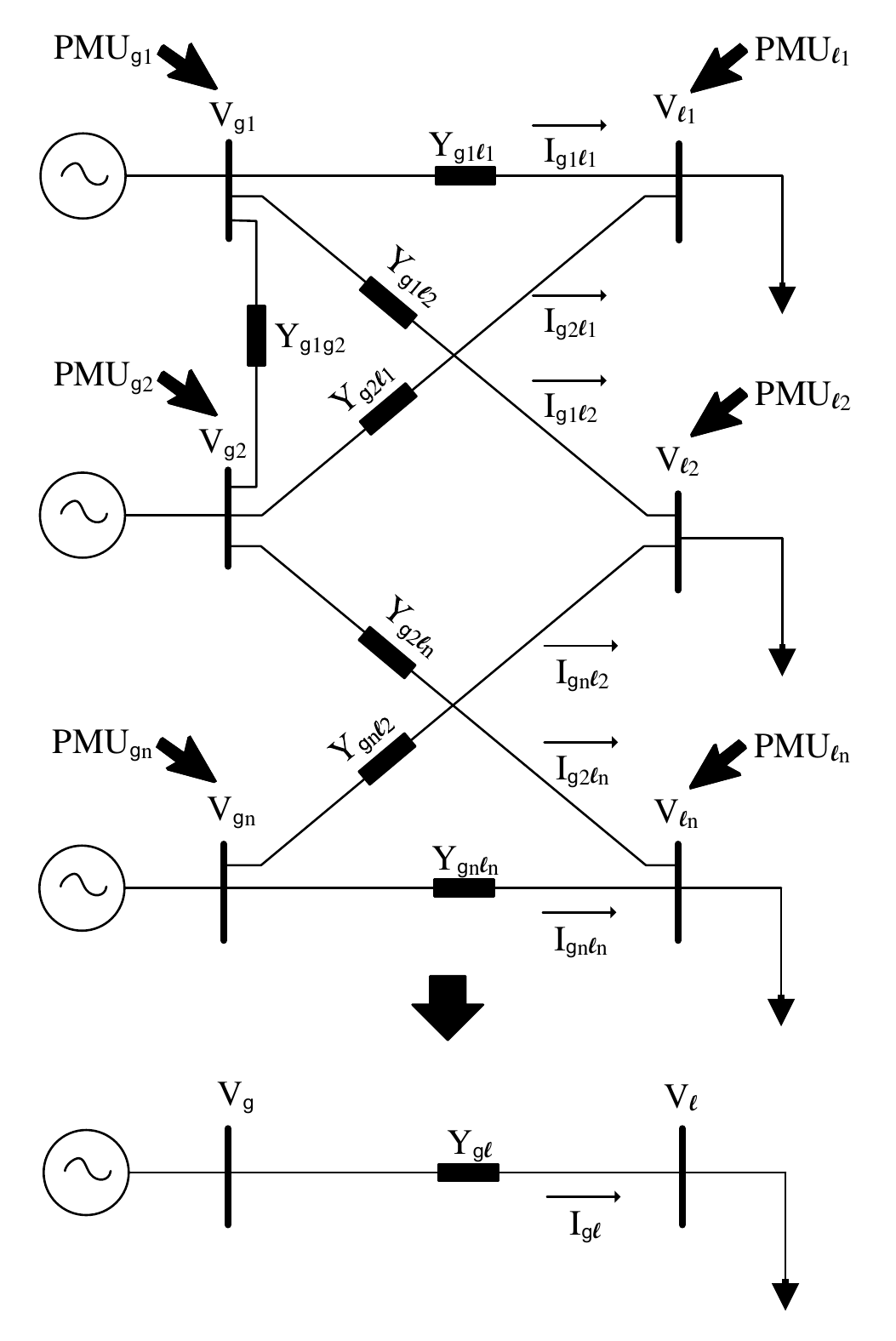}
\caption{
\label{Figure2corridor}%
Reduction of a power system with n-inputs and n-outputs to a single line system}%
\end{figure}

The reduction to a single line equivalent is done for a transmission corridor with $n$ inputs (generators at bus $g1$ to $gn$) and $n$ outputs (loads at buses $\ell1$ to $\ell n$), that will be reduced to an equivalent system with one input and one output, see Fig.\ref{Figure2corridor}.

The transmission corridor includes all the lines that are connecting the generation area with the load area. The generation area could have loads, but it has a net injection of power, and in the same way the load area could have generators but it is a net load.
The reduction of the corridor to an approximately equivalent single line enables the application of synchrophasor monitoring.
 
The complex currents and voltages ($I_{g1}$, ... $I_{gn}$, $I_{\ell1}$,... $I_{\ell n}$; $V_{g1}$,... $V_{gn}$, $V_{\ell1}$,... $V_{\ell n}$), are obtained from the PMUs at all the buses that bound the transmission corridor. Then the complex power is obtained from the measured complex currents and voltages:
\begin{align}
S_{gi}&=V_{gi}I_{gi}^*, &S_{\ell i}&= V_{\ell i}I_{\ell i}^*.
\end{align}
The  complete system will be reduced to a single line equivalent system while preserving the complex powers entering and leaving the corridors.
In other words, all the power that is entering (leaving) the transmission corridor is equal to all the power that is entering (leaving) the equivalent system. 
\begin{align}
\label{equalcomplexpowerg}
S_{g}&=\sum_{i=1}^{n}S_{gi},&S_{\ell}&=\sum_{i=1}^{n} S_{\ell i}.
\end{align}%
Similarly, the complex current entering (leaving) the transmission corridor is equal to the complex power entering (leaving) the equivalent system:
\begin{align}
I_{g}&=\sum_{i=1}^{n}I_{gi},& I_{\ell}&=\sum_{i=1}^{n} I_{\ell i}.
\label{equalcomplexcurrent}
\end{align}
Based the complex powers and complex currents of the transmission corridor and its equivalent, the voltages of the equivalent system are
\begin{align}
\label{vg}
V_{g}&=\frac{S_{g}}{{I_{g}}^*},&V_{\ell}&=\frac{S_{\ell}}{{I_{\ell}}^{*}}.
\end{align}
Then the voltage across, and admittance of the equivalent are: 
\begin{align}
\label{vgl}
V_{g\ell}&=\frac{S_{g}+S_{\ell}}{{I_{g}}^*},&
Y_{g\ell}&=V_{g\ell}I_{g}.
\end{align}
A benefit of this new reduction is that all the loads can change independently, making the model more realistic. In addition, we do not need to assume any admittance as known, which is very useful for online application, and for tracking the changes of the system such as contingencies.

\section{Methodology for evaluating voltage collapse margin across the transmission corridor with synchrophasors}

The methodology that we present in this section captures the effect of the outages in the transmission corridor through  the tracking of the voltage collapse margin. To obtain the voltage collapse margin across the transmission corridor, we require a PMU at both ends of the transmission lines that form the corridor. As the transmission corridors are composed of relatively few lines, we estimate that the number of PMUs required is between six and twenty, which is a feasible number of PMUs. 

In order to locate the bifurcation point of the system we assume a stable initial operating equilibrium and a slowly varying parameter, which is the increment of the load power varying slowly compared with the dynamics of the system. Under these assumptions, the power system can be modeled by static equations. In addition, to apply the usual synchrophasor monitoring approaches, the model is assumes PV  generation buses and PQ load buses. 

Generator reactive power limits are handled by sensing when generators reach their reactive power limits and changing that PV bus to a PQ bus in the transmission corridor model. That is, the generator with reactive power limits is modeled as a negative load.
The appropriate signals indicating generator reactive power limits can be obtained from standard control center signals or by processing PMU measurements at the generator.

The procedure is as follows:
\begin{enumerate}
\item	Measure with  PMUs the complex voltage and current at both ends of all the transmission lines that connect the generation area with the load area, see Fig.~\ref{Figure2corridor}.

\item	Check the reactive power limit signal of the generation buses. In case that the generator bus reaches its reactive limit then the bus is considered as PQ. For example, in this case we are considering that bus $g2$ of the system shown in Fig.~\ref{Figure2corridor} reaches its reactive limit.

\item Use the synchrophasor measurements to find the complex power of the generation area and the load area. The generation bus with reactive limits that changed to a PQ bus is treated as a negative load:
\begin{align}
S_{g}&=\sum_{i=1}^{n}S_{gi} - S_{g2}, &S_{\ell}&=\sum_{i=1}^{n} S_{\ell i}+S_{g2}.
\end{align}
\item Combine the complex current into the equivalent single line current:
\begin{align}
I_{g}&=\sum_{i=1}^{n} I_{gi}- I_{g2},&I_{\ell}&=\sum_{i=1}^{n} I_{\ell i}+I_{g2}.
\end{align}
\item Using the complex power of each area and the current  to find the voltages of the reduced system, see (\ref{vg}).

\item	Evaluate the voltage stability index from \cite{Milosevic03}:

\begin{align}
\label{VSIS}
\mbox{Index}&=\frac{{ \mid{V_{g\ell}}} \mid 100}{ \mid{V_{\ell}\mid}}.
\end{align}

\end{enumerate}
Index (\ref{VSIS}) indicates the maximum transfer of load that can be achieved across the transmission corridor that connect the areas under the measured condition of the corridor.  Using this index, an alarm can be triggered when a sufficient percentage of the index is exceeded.  For example, the alarm could be triggered when the index exceeds 80\%.

\section{Results}

\subsection{Evaluation of voltage collapse margin under multiple contingencies for WSCC 9-bus test system }

\begin{figure}[]
\centering
\includegraphics [width=0.5\textwidth]{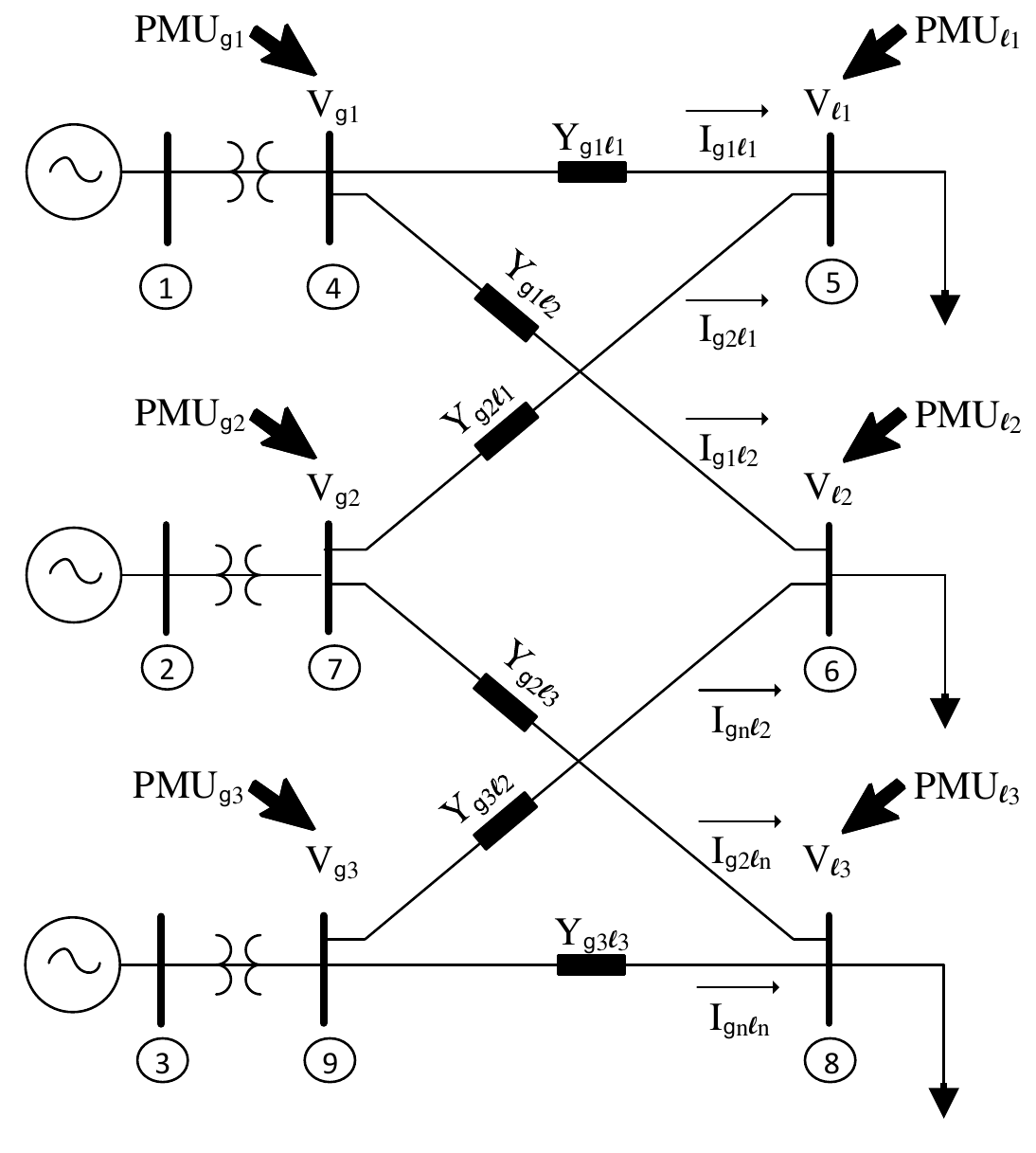}
\caption{
\label{9bus}%
WSCC 9-bus test system corridors with three inputs and three outputs}%
\end{figure}
In this section, we study the WSCC 9-bus test system shown in  Fig.~\ref{9bus}. We divide this system into two areas, the generation area and the load area. Each area is composed of three buses, and the areas are connected by six lines which form the transmission corridor.  This system  requires six PMUs to measure the complex voltage and current in all six buses bounding  the transmission corridor.

In order to explain better the reduction, we first evaluate the voltage stability margin of the   WSCC 9-bus test system without a contingency, see  Table \ref{Complete system}.  For this, we measure with the PMUs the complex voltage and currents in all the buses bounding the transmission corridor, and calculate the complex power entering or leaving at each bus. Combining the complex powers and current for the generation  and  load area, we obtain the equivalent voltage for each area, reducing the transmission corridor to a single line system to which the available methods for voltage stability in radial system can be applied.
\begin{table}[]
\centering
\caption{Reduction of WSCC 9-bus system and its voltage stability margin}
\label{Complete system}
\vspace{-2pt}
\begin{tabular}{lll}
 \hline\\[-5pt]
 \multicolumn{3}{c}{ PMU measurements}\\
$V_{g1}=1-j0.06$&$V_{g2}=1-j0.04$&$V_{g3}=1.00-j0.02$\\[1pt]
$V_{\ell1}=0.96-j0.12$&$V_{\ell2}=0.97-j0.1$&$V_{\ell3}=0.98-j0.07$\\[1pt]
$I_{g1}=1.17-j0.33$ &$I_{g2}=1-j0.08$&$I_{g3}=1+j0.12$\\[1pt]
$I_{\ell1}=-1.22+j0.67$&$I_{\ell2}=-0.88+j0.46$&$I_{\ell3}=-0.98+j0.48$\\[1pt]
\hline\\[-6pt]
\multicolumn{3}{c}{ Reduced System }\\
$V_{g}=1-j0.04$&$V_{\ell}=0.97-j0.1$&Index$=7\%$\\[1pt]
\hline\\[-8pt]
\end{tabular}
\end{table}

Now we simulate  n-1, n-2 and n-3 contingencies without shedding load to show the effect of contingencies on the voltage stability margin, see Table \ref{wescc}. 
In real time, when the contingency  occurs, the PMU measurements will track the changes in the voltage and current, then we use the methodology for reducing the corridor to a single line system and evaluate the voltage stability margin. This procedure is updating constantly to track the voltage stability margin across the transmission corridor.

The results shown in Table \ref{wescc} demonstrate that multiple outages can generate voltage stability problems. For example, under n-1 the highest voltage stability index is 28\%, however with an additional outage the voltage stability index can increase drastically to 85\%, indicating severe problems.

During  real operation, it is desirable to take remedial action promptly tracking the voltage stability margin, covering multiple contingencies in real time. For this, is necessary to define a security limit margin, which the operator should maintain in order to avoid voltage stability problems and blackout. In this way, if under any contingency the limit is violated, the operator should decrease the transfer of power across the corridor or in more severe cases shed load. For example, if we consider the security margin as eighty percent, under contingency number 20, the system operator should take action in order to maintain the security level required, see Table \ref{wescc}.

Additionally, in Table \ref{wescc}, we evaluate the accuracy of this approach. We contrast the voltage stability index using the synchrophasor measurements with the exact answers for the stability margin obtained using the well known continuation power flow. These results shown that our method for combining multiple lines using measurements is a reasonable approximation, with an error in the index less than 15\%. The error reduces for the more highly stressed cases of interest. For future work, we will analyze and explain the origin of the error. 

\begin{table}[]
\centering
\caption{Voltage collapse margin under contingencies for WSCC 9-bus system}
\label{wescc}
\vspace{-2pt}
\begin{tabular}{cccccc}
\multicolumn{1}{c}{Contingency}& \multicolumn{3}{c}{Line outages}& \multicolumn{2}{c}{Voltage stability margin (\%)}\\[2pt]
&&&& PMUs&Continuation PF\\[2pt]
 \hline\\[-5pt]
1&&&&7&22\\[1pt]
2&4-6&&&10&24\\[1pt]
3&8-9&&&10&24\\[1pt]
4&5-7&&&10&24\\[1pt]
5&4-5&&&17&30\\[1pt]
6&7-8&9.6&&20&32\\[1pt]
7&7-8&&&21&34\\[1pt]
8&4-6&7-5&&21&35\\[1pt]
9&7-5&9-8&&23&36\\[1pt]
10&4-5&9-8&&24&36\\[1pt]
11&6-9&&&28&40\\[1pt]
12&7-5&7-8&&31&46\\[1pt]
13&4-6&7-8&&37&50\\[1pt]
14&4-5&7-8&&40&52\\[1pt]
15&4-5&4-6&&52&62\\[1pt]
16&4-5&9-6&&60&69\\[1pt]
17&4-6&9-8&&60&69\\[1pt]
18&4-6&9-8&7-5&61&70\\[1pt]
19&4-6&7-8&7-5&70&75\\[1pt]
20&7-5&9-6&&85&89\\
 \hline\\[-6pt]
\end{tabular}
\end{table}

\section{Conclusion}
\label{conclusion}

We show how to reduce  multiple lines in several transmission corridors to a single line equivalent to 
which online monitoring of voltage stability with synchrophasors can be applied.  The reduction is based on synchrophasor measurements 
of complex power and current at both ends of the lines, and the reduction shows how to 
combine the synchrophasor measurements so that they are effective in monitoring voltage stability.

The approach can give a fast, online  indication of voltage stability that can accommodate both
multiple contingencies and generator reactive power limits.
These capabilities should increase operator situational awareness under emergency conditions, and should be 
complementary to methods that make pre-contingency calculations from a model based on the state estimator.
 
 The new methodology for analyzing online voltage stability margin for multiple transmission lines and multiple contingencies was tested in the WSCC 9-bus  system. Our results suggest that we have found a promising and systematic approach for online monitoring of voltage stability margin for multiple transmission lines and multiple contingencies. Planned future work will generalize the methodology for corridors that include load or generation inside the corridors and analyze the approximations made.

\vspace{10pt}

\section{Acknowledgements}
\label{ack}
We gratefully acknowledge support in part from the Electric Power Research Center at Iowa State University, 
Arend J. and Velma V. Sandbulte professorship funds, and
NSF grant CPS-1135825.

\vspace{10pt}

\bibliographystyle{./IEEEtran}
\bibliography{./IEEEabrv,./Paper1ref}

\begin{thebibliography}{10}
\providecommand{\url}[1]{#1}
\csname url@samestyle\endcsname
\providecommand{\newblock}{\relax}
\providecommand{\bibinfo}[2]{#2}
\providecommand{\BIBentrySTDinterwordspacing}{\spaceskip=0pt\relax}
\providecommand{\BIBentryALTinterwordstretchfactor}{4}
\providecommand{\BIBentryALTinterwordspacing}{\spaceskip=\fontdimen2\font plus
\BIBentryALTinterwordstretchfactor\fontdimen3\font minus
  \fontdimen4\font\relax}
\providecommand{\BIBforeignlanguage}[2]{{%
\expandafter\ifx\csname l@#1\endcsname\relax
\typeout{** WARNING: IEEEtran.bst: No hyphenation pattern has been}%
\typeout{** loaded for the language `#1'. Using the pattern for}%
\typeout{** the default language instead.}%
\else
\language=\csname l@#1\endcsname
\fi
#2}}
\providecommand{\BIBdecl}{\relax}
\BIBdecl

\bibitem{5528820}
K.~Nara, K.~Tanaka, H.~Kodama, R.~Shoults, M.-S. Chen, P.~Van~Olinda, and
  D.~Bertagnolli, ``On-line contingency selection algorithm for voltage
  security analysis,'' \emph{IEEE Power Engineering Review}, vol. PER-5, no.~4,
  pp. 41--42, April 1985.

\bibitem{EjebePS88}
G.~Ejebe, H.~Van~Meeteren, and B.~Wollenberg, ``Fast contingency screening and
  evaluation for voltage security analysis,'' \emph{IEEE Trans. Power Systems},
  vol.~3, no.~4, pp. 1582--1590, Nov 1988.

\bibitem{HadjsaidPS93}
N.~Hadjsaid, M.~Benahmed, J.~Fandino, J.~Sabonnadiere, and G.~Nerin, ``Fast
  contingency screening for voltage-reactive considerations in security
  analysis,'' \emph{IEEE Trans. Power Systems}, vol.~8, no.~1, pp. 144--151,
  Feb 1993.

\bibitem{515191}
G.~Ejebe, G.~Irisarri, S.~Mokhtari, O.~Obadina, P.~Ristanovic, and J.~Tong,
  ``Methods for contingency screening and ranking for voltage stability
  analysis of power systems,'' in \emph{Power Industry Computer Application
  Conference}, May 1995, pp. 249--255.

\bibitem{387897}
H.-D. Chiang, A.~Flueck, K.~Shah, and N.~Balu, ``Cpflow: a practical tool for
  tracing power system steady-state stationary behavior due to load and
  generation variations,'' \emph{IEEE Trans. Power Systems}, vol.~10, no.~2,
  pp. 623--634, May 1995.

\bibitem{486117}
G.~Ejebe, G.~Irisarri, S.~Mokhtari, O.~Obadina, P.~Ristanovic, and J.~Tong,
  ``Methods for contingency screening and ranking for voltage stability
  analysis of power systems,'' \emph{IEEE Trans. Power Systems}, vol.~11,
  no.~1, pp. 350--356, Feb 1996.

\bibitem{ChiangPS97}
H.-D. Chiang, C.-S. Wang, and A.~Flueck, ``Look-ahead voltage and load margin
  contingency selection functions for large-scale power systems,'' \emph{IEEE
  Trans. Power Systems}, vol.~12, no.~1, pp. 173--180, Feb 1997.

\bibitem{826465}
S.~Repo and P.~Jarentausta, ``Contingency analysis for a large number of
  voltage stability studies,'' in \emph{International Conference on Electric
  Power Engineering, PowerTech Budapest}, Aug 1999, pp. 34--.

\bibitem{GreenePS99}
S.~Greene, I.~Dobson, and F.~Alvarado, ``Contingency ranking for voltage
  collapse via sensitivities from a single nose curve,'' \emph{IEEE Trans.
  Power Systems}, vol.~14, no.~1, pp. 232--240, Feb 1999.

\bibitem{FlueckPS00}
A.~Flueck and J.~Dondeti, ``A new continuation power flow tool for
  investigating the nonlinear effects of transmission branch parameter
  variations,'' \emph{IEEE Trans. Power Systems}, vol.~15, no.~1, pp. 223--227,
  Feb 2000.

\bibitem{1033089}
I.~Musirin, T.~Khawa, and A.~Rahman, ``Simulation technique for voltage
  collapse prediction and contingency ranking in power system,'' in
  \emph{Student Conference on Research and Development}, 2002, pp. 188--191.

\bibitem{FlueckPS02}
A.~Flueck, R.~Gonella, and J.~Dondeti, ``A new power sensitivity method of
  ranking branch outage contingencies for voltage collapse,'' \emph{IEEE Trans.
  Power Systems}, vol.~17, no.~2, pp. 265--270, May 2002.

\bibitem{1397706}
M.~Poshtan, P.~Rastgoufard, and B.~Singh, ``Contingency ranking for voltage
  stability analysis of large-scale power systems,'' in \emph{IEEE PES Power
  Systems Conference and Exposition}, Oct 2004, pp. 1506--1513 vol.3.

\bibitem{FlueckPS04}
W.~Qiu and A.~Flueck, ``A new technique for evaluating the severity of
  generator outage contingencies based on two-parameter continuation,'' in
  \emph{IEEE PES Power Systems Conference and Exposition}, Oct 2004, pp.
  150--156 vol.1.

\bibitem{1425595}
N.~Amjady and M.~Esmaili, ``Application of a new sensitivity analysis framework
  for voltage contingency ranking,'' \emph{IEEE Trans. Power Systems}, vol.~20,
  no.~2, pp. 973--983, May 2005.

\bibitem{4410569}
A.~Tiwari and V.~Ajjarapu, ``Contingency assessment for voltage dip and short
  term voltage stability analysis,'' in \emph{IREP Symposium, Bulk Power System
  Dynamics and Control - VII}, Aug 2007, pp. 1--8.

\bibitem{4809089}
T.~Srinivas, K.~Reddy, and V.~Devi, ``Composite criteria based network
  contingency ranking using fuzzy logic approach,'' in \emph{IEEE International
  Advance Computing Conference}, March 2009, pp. 654--657.

\bibitem{5442743}
C.~Subramani, S.~Dash, M.~Arun~Bhaskar, M.~Jagadeeshkumar, K.~Sureshkumar, and
  R.~Parthipan, ``Line outage contingency screening and ranking for voltage
  stability assessment,'' in \emph{International Conference on Power Systems},
  Dec 2009, pp. 1--5.

\bibitem{5376473}
C.~Subramani, S.~Dash, M.~Bhaskar, M.~Jagadeeshkumar, and K.~Balaji, ``Soft
  computing for voltage stability analysis and contingency ranking in power
  system,'' in \emph{International Conference on Advances in Computing,
  Control, Telecommunication Technologies}, Dec 2009, pp. 583--587.

\bibitem{6156658}
G.~Kumar and M.~Kalavathi, ``Cpf, tds based voltage stability analysis using
  series, shunt and series-shunt facts controllers for line outage
  contingency,'' in \emph{International Conference on Power and Energy
  Systems}, Dec 2011, pp. 1--6.

\bibitem{Varshney13}
S.~Varshney, L.~Srivastava, M.~Pandit, and M.~Sharma, ``Voltage stability based
  contingency ranking using distributed computing environment,'' in
  \emph{International Conference on Power, Energy and Control}, Feb 2013, pp.
  208--212.

\bibitem{Glavic11}
M.~Glavic and T.~Van~Cutsem, ``A short survey of methods for voltage
  instability detection,'' in \emph{IEEE PES General Meeting}, July 2011, pp.
  1--8.

\bibitem{RamirezGM14}
L.~Ramirez and I.~Dobson, ``Monitoring voltage collapse margin by measuring the
  area voltage across several transmission lines with synchrophasors,'' in
  \emph{IEEE PES General Meeting}, July 2014, pp. 1--5.

\bibitem{Chakrabortty11}
A.~Chakrabortty, J.~Chow, and A.~Salazar, ``A measurement-based framework for
  dynamic equivalencing of large power systems using wide-area phasor
  measurements,'' \emph{IEEE Transactions on Smart Grid}, vol.~2, no.~1, pp.
  68--81, March 2011.

\bibitem{WarlandPSC02}
L.~Warland and A.~T. Holen, ``Estimation of distance to voltage collapse:
  Testing an algorithm based on local measurements,'' \emph{14th Power Systems
  Computation Conference, Sevilla}, 2002.

\bibitem{LarssonBPT03}
M.~Larsson, C.~Rehtanz, and J.~Bertsch, ``Monitoring and operation of
  transmission corridors,'' in \emph{IEEE Power Tech Conference Proceedings,
  Bologna}, vol.~3, June 2003, pp. 8 pp. Vol.3--.

\bibitem{Mingsong08}
M.~Liu, B.~Zhang, L.~Yao, M.~Han, H.~Sun, and W.~Wu, ``Pmu based voltage
  stability analysis for transmission corridors,'' in \emph{Third International
  Conference on Electric Utility Deregulation and Restructuring and Power
  Technologies}, April 2008, pp. 1815--1820.

\bibitem{Milosevic03}
B.~Milosevic and M.~Begovic, ``Voltage-stability protection and control using a
  wide-area network of phasor measurements,'' \emph{IEEE Trans. Power Systems},
  vol.~18, no.~1, pp. 121--127, Feb 2003.

\bibitem{DuongEPECS13}
D.~Duong and K.~Uhlen, ``Online voltage stability monitoring based on pmu
  measurements and system topology,'' in \emph{International Conference on
  Electric Power and Energy Conversion Systems}, Oct 2013, pp. 1--6.
\renewcommand{\BIBentryALTinterwordstretchfactor}{4}

\end{thebibliography}

\end{document}